\documentclass[twocolumn,times]{aastex61} 

\usepackage{siunitx} 

\received{2017 December 22}
\revised{2018 February 7}
\accepted{2018 February 21}

\submitjournal{ApJL} 

\shorttitle{A Model of a Flux Rope that Formed in the Solar Corona}
\shortauthors{James et al.}

\begin{document}

\title{An Observationally-Constrained Model of a Flux Rope that Formed in the Solar Corona}

\correspondingauthor{Alexander W. James}
\email{alexander.james.15@ucl.ac.uk}

\author[0000-0001-7927-9291]{Alexander W. James}
\affil{Mullard Space Science Laboratory, University College London Holmbury St. Mary, Dorking Surrey, RH5 6NT, UK}

\author[0000-0001-7809-0067]{Gherardo Valori}
\affil{Mullard Space Science Laboratory, University College London Holmbury St. Mary, Dorking Surrey, RH5 6NT, UK}

\author[0000-0002-0053-4876]{Lucie M. Green}
\affil{Mullard Space Science Laboratory, University College London Holmbury St. Mary, Dorking Surrey, RH5 6NT, UK}

\author[0000-0002-0671-689X]{Yang Liu}
\affiliation{W.W. Hansen Experimental Physics Laboratory, Stanford University, Stanford, CA 94305-4085, USA}

\author[0000-0003-2110-9753]{Mark C. M. Cheung}
\affiliation{Lockheed Martin Solar and Astrophysics Laboratory 3521 Hanover Street Bldg. 252, Palo Alto, CA 94304, USA}

\author[0000-0002-9293-8439]{Yang Guo}
\affiliation{School of Astronomy and Space Science and Key Laboratory of Modern Astronomy and Astrophysics in Ministry of Education,
Nanjing University, Nanjing 210023, China}

\author[0000-0002-2943-5978]{Lidia van Driel-Gesztelyi}
\affiliation{Mullard Space Science Laboratory, University College London Holmbury St. Mary, Dorking Surrey, RH5 6NT, UK}
\affiliation{Observatoire de Paris, LESIA, FRE 2461 (CNRS), 92195 Meudon Principal Cedex, France}
\affiliation{Konkoly Observatory of the Hungarian Academy of Sciences, Budapest, Hungary}

\begin{abstract} 
Coronal mass ejections (CMEs) are large-scale eruptions of plasma from the coronae of stars.
Understanding the plasma processes involved in CME initiation has applications to space weather forecasting and laboratory plasma experiments.
\citeauthor{james2017on-disc} (Sol. Phys. 292, 71, 2017) used EUV observations to conclude that a magnetic flux rope formed in the solar corona above NOAA Active Region 11504 before it erupted on 14 June 2012 (SOL2012-06-14).
In this work, we use data from the Solar Dynamics Observatory to model the coronal magnetic field of the active region one hour prior to eruption using a nonlinear force-free field extrapolation, and find a flux rope reaching a maximum height of 150 Mm above the photosphere. 
Estimations of the average twist of the strongly asymmetric extrapolated flux rope are between 1.35 and 1.88 turns, depending on the choice of axis, although the erupting structure was not observed to kink.
The decay index near the apex of the axis of the extrapolated flux rope is comparable to typical critical values required for the onset of the torus instability, so we suggest that the torus instability drove the eruption.
\end{abstract}



\section{Introduction} \label{sec:intro}

Coronal mass ejections (CMEs) are large-scale eruptions of plasma from the coronae of stars
, and there is currently no consensus regarding their cause.
Understanding the stability of the plasma structures involved in CME initiation is important for forecasting space weather sufficiently far in advance of CME arrival at Earth.

Solar CMEs have large kinetic energies ($\sim$\num{e32} ergs; \citealt{forbes2000review}), so many CME models describe processes by which an increasing amount of energy is stored in the corona and then suddenly released \citep{forbes2000review}.
The question of what processes cause this storage and release of energy in CMEs is closely tied to understanding the pre-eruptive configuration of the corona, because certain eruption mechanisms may only be relevant to specific topologies.
Measurements taken at 1 AU have in some cases revealed that CMEs arriving at Earth contain twisted magnetic structures called flux ropes \citep{burlaga1981magnetic}, however it is debated whether these flux ropes form before or after CME onset \citep{antiochos1999model,moore2001onset,green2009flux}.

There are a wide range of proposed CME triggers that bring the coronal magnetic field to the brink of eruption, including rotation of sunspot fragments around each other (\textit{e.g.}, \citealt{yan2012sunspot,james2017on-disc}), magnetic reconnection (\textit{e.g.}, flux cancellation; \citealt{vanBallegooijen1989formation}, and tether-cutting; \citealt{moore2001onset}), and the helical kink instability of a flux rope \citep{torok2005kink}, but there are only two main groups of theories regarding the process that drives the rapid expansion of CMEs.
One group assumes that CMEs are driven by an ideal magnetohydrodynamic instability involving a flux rope, such as the torus instability \citep{vanTend1978development,kliem2006torus,demoulin2010criteria}, whereas the other set of theories assumes that CMEs are driven by flare reconnection \citep{antiochos1999model,temmer2010combined,karpen2012mechanisms}. 

\citeauthor{james2017on-disc} (\citeyear{james2017on-disc}; hereafter Paper I) observationally studied the pre-eruptive configuration of a CME that occurred on 14 June 2012.
The CME originated from NOAA Active Region 11504 when it was near the centre of the solar disc.
EUV images show a transient sigmoid during a confined flare two hours before the CME, suggesting the presence of a flux rope before the onset of eruption, and the flux rope footpoints were inferred by EUV dimmings and flare ribbons. 
For details on observational signatures of flux ropes, see Paper I and references within.
Observations from a number of EUV channels suggest the flux rope formed by reconnection in the corona rather than in the photosphere or chromosphere, which is confirmed by the measured coronal plasma composition in the flux rope.
For more examples of flux ropes formed via coronal reconnection, see \citet{patsourakos2013direct,nindos2015common}.

The conclusions of Paper I were based on indirect indications of the flux rope, because we are presently unable to directly measure the coronal magnetic field.
However, techniques have been developed to extrapolate the coronal magnetic field from complex (but routinely available) photospheric measurements under the assumption that the corona is in a force-free state (see \citealt{wiegelmann2012force-free}), \textit{i.e.},
 \begin{equation}
 \textbf{J} \times \textbf{B} = 0,
 \label{eqn:curlB}
 \end{equation}
where \textbf{B} is the magnetic field vector and \textbf{J} is the electric current density:
 \begin{equation}
 \textbf{J} = \frac{1}{\mu_{0}} \nabla \times \textbf{B}.
 \label{eqn:J}
 \end{equation}
 
This assumption is satisfied by either a potential field ($\textbf{J} = 0$), or a field in which electric currents are parallel to the magnetic field vector ($\nabla \times \textbf{B} = \alpha \textbf{B}$).
In the nonlinear case, the force-free parameter, $\alpha$, varies for different field lines.
The nonlinear force-free field (NLFFF) approximation is the simplest method that can reproduce the electric currents and complex distributions of twist associated with a flux rope embedded in an arcade.

In this work, we use a NLFFF extrapolation to test the hypothesis of Paper I that a flux rope formed before the CME that occurred at $\approx$13:30 UT on 14 June 2012, and investigate the cause of its eruption. 

\section{Data, Method, and Validation of the Model} \label{sec:data}

The NLFFF extrapolation of NOAA Active Region 11504 was performed using a photospheric magnetogram produced by the \textit{Helioseismic and Magnetic Imager} (\textit{HMI}; \citealt{scherrer2012hmi}) onboard the \textit{Solar Dynamics Observatory} (\textit{SDO}; \citealt{pesnell2012SDO}). 
The chosen magnetogram was taken at $\approx$12:24 UT on 14 June 2012: approximately one hour after the first observational indication that a flux rope was present, and one hour before the CME began (see Paper I).

 \begin{figure*}[!htb]
 \centerline{\includegraphics[width=1.0\textwidth,clip=]{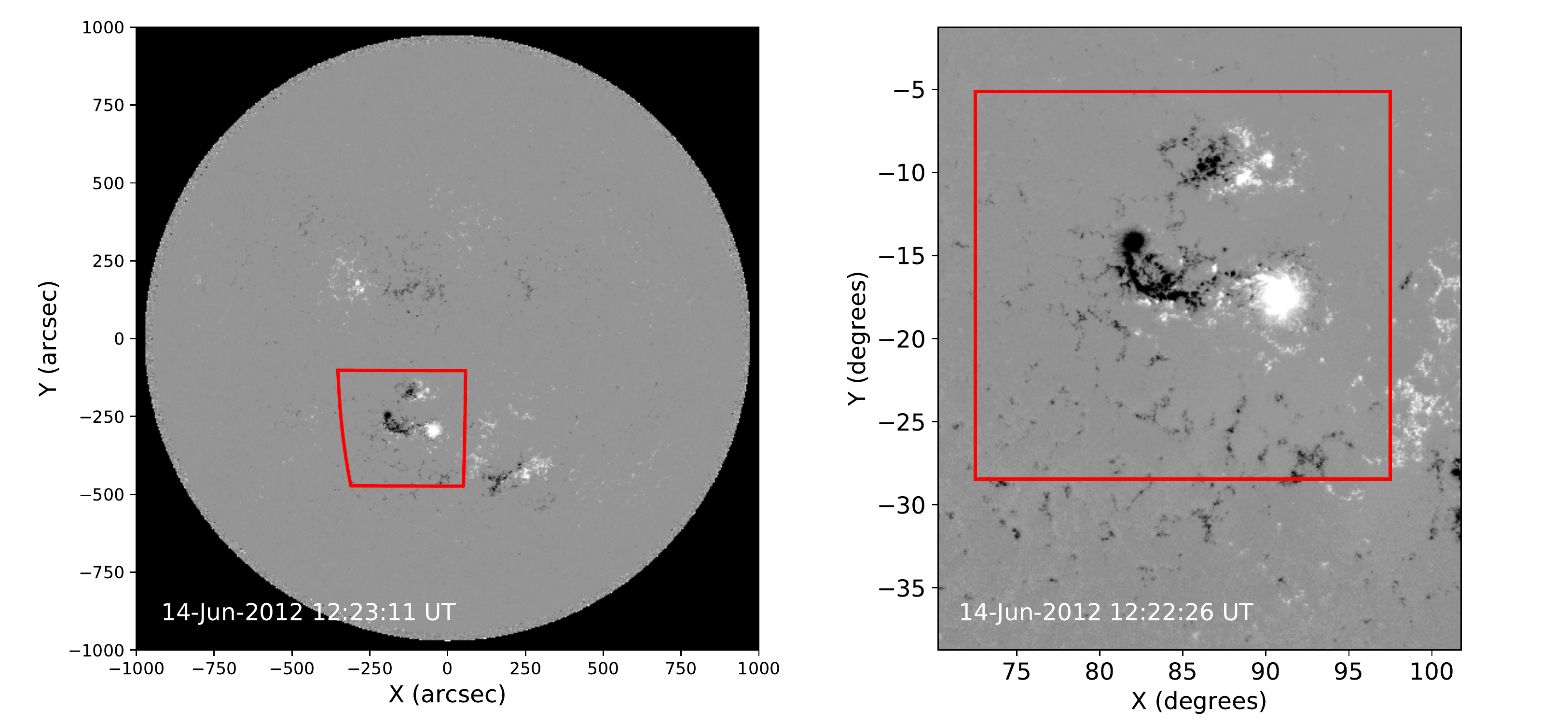}}
 \caption{\textit{HMI} magnetograms of the Sun taken on 14 June 2012. Positive (negative) magnetic flux is shown in white (black), and saturated at $\pm$1000 G. The left panel is a line-of-sight magnetogram for context, and the right panel shows the radial magnetic field component in cylindrical equal-area projection. The red boxes show the boundary of the SHARP-style magnetogram that was used for the NLFFF extrapolation, containing NOAA Active Region 11504.}
 \label{fig:HMI}
 \end{figure*}
 
 \begin{figure*}[!htb]
 \centerline{\includegraphics[width=1.0\textwidth,clip=]{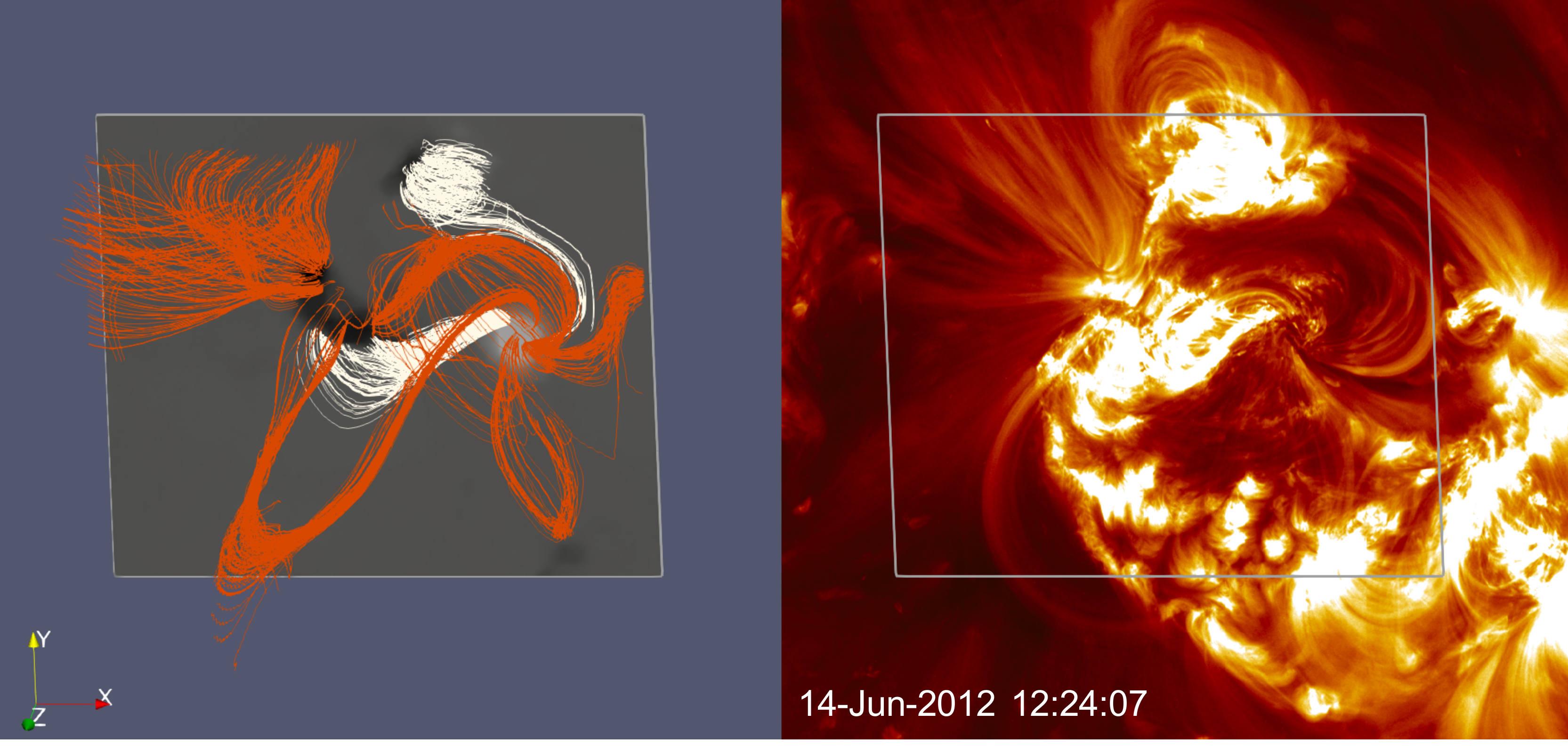}}
 \caption{The extrapolated coronal magnetic field (left) closely matches a number of active region features observed in the 193 \AA{} channel of \textit{AIA} (right). The \textit{AIA} image is saturated at 2500 DN s$^{-1}$ pixel$^{-1}$.}
 \label{fig:euv_comparison}
 \end{figure*}
 
The observed EUV sigmoid extended to the south of the NOAA Active Region 11504 before the CME occurred (see Figure 5c of Paper I), and so the field-of-view of the publicly available \textit{HMI} SHARP series magnetogram \citep{bobra2014sharp} was too small to accurately reproduce the sigmoidal field in the extrapolation.
A bespoke SHARP-style magnetogram was produced with a large enough field-of-view to accommodate the size of the sigmoid ($\approx 400''\times600''$), whilst still excluding as much of the dispersed negative magnetic flux to the south of the active region as possible (see the red boxes in Figure \ref{fig:HMI}).

The magnetogram was re-binned to 1/6th resolution, such that each pixel in the cylindrical equal-area projection represents an angular diameter of 0.18$^{\circ}$ (equivalent to 3$''$ or $\approx$2.18 Mm at disc centre when viewed from 1 AU) and smoothed using the median of a 7-pixel boxcar.
To alleviate the impact on the extrapolated field caused by the isolated negative magnetic flux to the south of the active region, balance of positive and negative magnetic flux was enforced over the magnetogram, corresponding to an increase of 5.2 G in each pixel (which is far smaller than the 100 G error estimation suggested by \citealt{hoeksema2014helioseismic} and corresponds to a total change of 13.6\% in the open field flux).
The magnetogram was then pre-processed using the method of \citet{fuhrmann2007preprocessing} to reduce the total Lorentz force by applying variations to the horizontal and vertical magnetic field components.
Modifications to the observed horizontal (vertical) field component in each pixel were limited to 80 G or 30$\%$ (30 G or 10$\%$) of its initial value - whichever is largest.

The coronal magnetic field was then extrapolated from the pre-processed magnetogram using the magnetofrictional NLFFF method detailed by \citet{valori2010testing}, yielding a model of the AR with force-free parameter $\sigma_{J}$$\approx$25$\%$ \citep{wheatland2000optimization} and solenoidal error limited to 9$\%$ of the total energy \citep{valori2013accuracy}.
The top boundary of the extrapolation volume was chosen to sufficiently accommodate the height of a toroidal flux rope with a footpoint separation as indicated in Figure 6 of Paper I.

To check the validity of the extrapolation, we compare the extrapolated magnetic field to EUV observations of the active region produced by the \textit{Atmospheric Imaging Assembly} (\textit{AIA}; \citealt{lemen2012aia}) onboard \textit{SDO}
after re-projecting the NLFFF extrapolated field lines onto the \textit{AIA} plane-of-sky images (as in \citealt{polito2017analysis}).
Figure \ref{fig:euv_comparison} shows that the extrapolation reproduces the large-scale active region emission structures seen in the 193 \AA{} channel of \textit{AIA}, including field lines that fan out from the edges of the active region and a sheared arcade in the core of the active region (shown by the southern group of white field lines in Figure \ref{fig:euv_comparison}a) that matches the observed sheared arcade (see Section 3.3 and Figure 5 of Paper I for more details).

\section{The Pre-Eruptive Flux Rope} \label{sec:analysis}


 \begin{figure*}[!htb]
 \centerline{\includegraphics[width=1.0\textwidth,clip=]{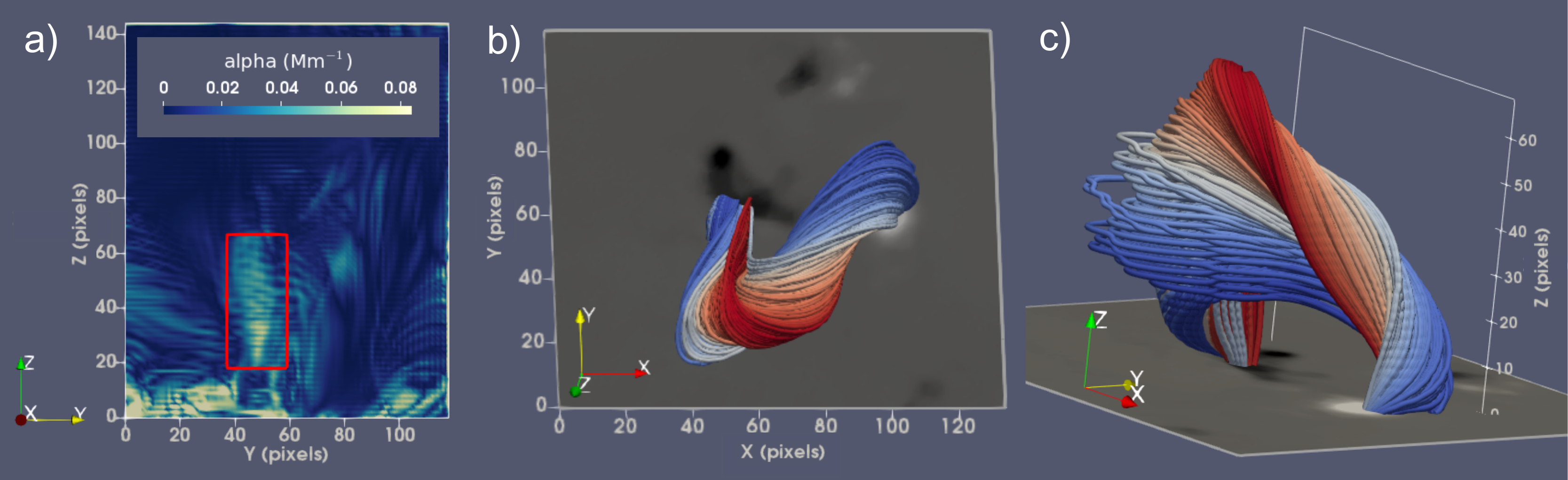}}
 \caption{a) Vertical slice at x=67 pixels to show $\alpha$ through the extrapolation volume. A large region of high $\alpha$ is outlined by a red box and used to define the extent of the flux rope. b) The flux rope in the extrapolated field as seen from the same perspective as \textit{SDO}. c) Side-on view of the extrapolated flux rope. Flux rope stream lines are drawn through the selected region of $\alpha>$0.02 Mm$^{-1}$ shown in panel a and are each given a fixed colour along their length.}
 \label{fig:flux_rope_alpha}
 \end{figure*}

We use the force-free parameter, $\alpha$, as a proxy for twist in the extrapolated magnetic field.
Figure \ref{fig:flux_rope_alpha}a shows a volume between the active region sunspots in the NLFFF model where $\alpha$ is large relative to the rest of the extrapolated field (see Figure \ref{fig:flux_rope_alpha}a).
Visualising magnetic field lines that pass through this region where $\alpha >$ 0.02 Mm$^{-1}$ reveals a flux rope (see Figure \ref{fig:flux_rope_alpha}b and \ref{fig:flux_rope_alpha}c).
The location and shape of the flux rope remarkably matches most, if not all, of the observational constraints identified in Section 4 of Paper I.

The flux rope extends high in the corona, with its highest point reaching $\approx$150 Mm ($\approx$$0.2R_{\odot}$) above the photosphere.
The axis of the flux rope is not planar, but is oriented roughly eastward and inclined to the south with respect to the vertical. 
The flux rope is highly asymmetric and has a strongly inhomogeneous distribution of right-handed twist.
The flux rope cross-section shown in Figure \ref{fig:flux_rope_alpha}a has a major diameter of $\approx$105 Mm and a minor diameter of $\approx$35 Mm.

The footpoints of the extrapolated flux rope are located in the north-western penumbra of the positive sunspot and to the south of the negative sunspot.
EUV observations, however, suggest that the western footpoint of the flux rope was rooted in the south-western penumbra of the positive sunspot during the eruption (see Figure 6 of Paper I).
This difference of approximately half the sunspot diameter could be due to modification of the magnetogram induced by preprocessing, to the difference in time between the pre-eruptive extrapolation and observations during the dynamic phase of the eruption, or to projection effects in the coronal EUV data.

 \begin{figure*}[!htb]
 \centerline{\includegraphics[width=0.7\textwidth,clip=]{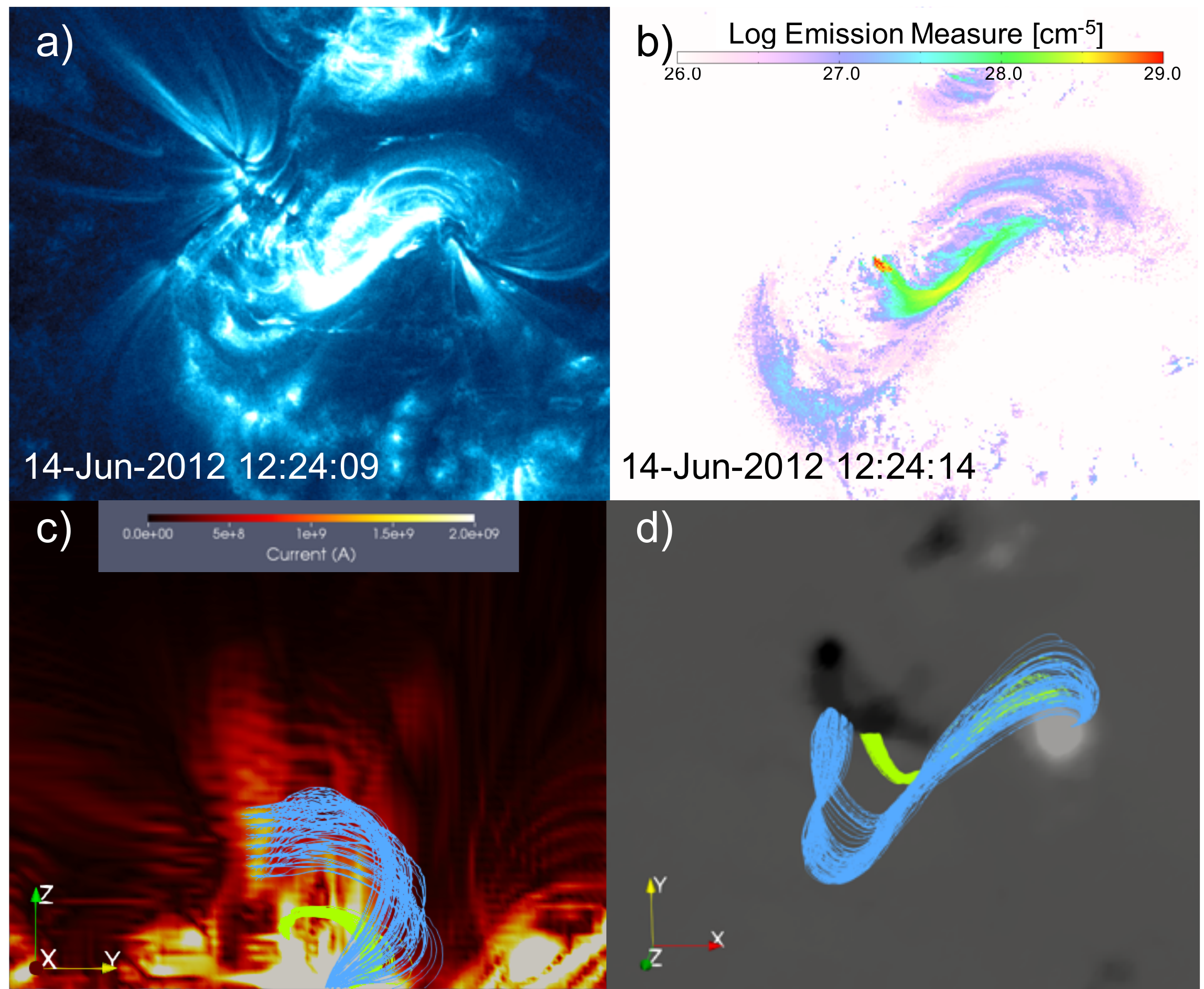}}
 \caption{a) EUV image of NOAA Active Region 11504 in the 131 \AA{} channel of \textit{AIA} showing the bright central flare arcade and the faint sigmoid. The image is saturated at $\pm$200 DN s$^{-1}$ pixel$^{-1}$. b) Differential emission measure of NOAA Active Region 11504 in the temperature range log(T/K)=6.85--7.15 shows the flare arcade and sigmoid. c) Electric current in a vertical slice taken through the extrapolation volume. The flux rope appears as a region of relatively high current, which is particularly strong at the bottom of the flux rope. Stream lines are drawn through regions of strong current that match the EUV observations of the sigmoid (blue) and the flare arcade (green). d) The extrapolated sigmoidal and flare arcade stream lines from panel c as viewed from the perspective of \textit{SDO}, imposed on the extrapolated magnetogram.}
 \label{fig:dem_comparison}
 \end{figure*}

In addition to the set of observations presented in Paper I, a differential emission measure (DEM) inversion was performed using the method of \citet{cheung2015thermal} to study the thermal emission of the active region. 
At the time of the extrapolation, the sigmoid and underlying flare arcade observed in the active region emit most strongly in the temperature range log(T/K)=6.85--7.15.
This is consistent with their observation in the 131 \AA{} channel of \textit{AIA}, which has a peak in temperature response at 11 MK \citep{lemen2012aia}, and confirms flux rope temperatures from previous DEM studies (\textit{e.g.}, \citealt{cheng2012differential}).
The shape of the sigmoid in the EUV observations and DEM closely matches extrapolated field lines that pass through the strong region of current density in the bottom-third of the flux rope (see Figure \ref{fig:dem_comparison}).

Field lines that reproduce the observed sheared arcade were also found to pass through a region of strong current beneath the flux rope (see the green field lines in Figure 4c and 4d; the average value of $\alpha$ here is $\approx$0.07 Mm$^{-1}$).
In addition to the many observational details summarised above, the similarly hot temperatures and high current densities of the sigmoid and the arcade support the hypothesis of Paper I that the flux rope and flare arcade form as the products of magnetic reconnection in the corona.

The flux rope (as defined by the region of large $\alpha$ in the red box in Figure \ref{fig:flux_rope_alpha}a) contains $\num{4e20}$ Mx of magnetic flux ($\approx$3$\%$ of half the unsigned active region flux).
The average value of $\alpha$ in the flux rope is $\approx$0.07 Mm$^{-1}$, as in the part of the sheared arcade where the current density is stronger.
The total electric current within the flux rope is \num{2.3e11} A, with an average current density of {\num{9.9e-5} A m$^{-2}$. 
These values are similar in magnitude to previous estimations of currents in prominences (
see \citealt{Filippov2015solar} and references within).

\subsection{Twist and Writhe} \label{sec:twist}

The flux rope is twisted and extends very high in the atmosphere, so we investigate whether it is actually stable.
According to the test in Section 4.3 of \citet{valori2010testing}, an unstable equilibrium would result in an uncommonly long computational time, which was not recorded in this case.

The helical kink instability will occur if the flux rope twist exceeds a critical value.
To quantify the twist in our very asymmetric case, the twist of individual field lines in the flux rope was calculated around an axis, and the average was taken \citep{guo2010driving,guo2013twist}.
Following \citet{guo2017magnetic}, the axial field line is defined as the field line with the smallest ratio of tangential-to-normal magnetic field components with respect to a plane roughly perpendicular to the body of the flux rope. 

Given the marked asymmetry of the flux rope, we took three slices through the flux rope with different inclinations, resulting in three axes and therefore three values of the average twist, namely 1.35, 1.61, and 1.88 turns.
The axes are similar in height and length (for reference, the first axis is $\approx$350 Mm long and reaches up to $\approx$120 Mm above the photosphere).
The same set of flux rope field lines were used to determine the twist around each of the three axes. 

\citet{torok2003twisting} examine a \citet{titov1999topology} flux rope topology and find a critical twist of 3.5$\pi$ (1.75 turns) for the onset of the helical kink instability.
This twist threshold is comparable to two of the values obtained for the asymmetric, non-uniform flux rope in this work.

The observations detailed in Paper I show no significant sign of the flux rope kinking before or during the eruption. 
In fact, the axis of the flux rope changed so little that the CME configuration measured \textit{in situ} closely matched the pre-eruptive configuration \citep{palmerio2017insitu}. 
Therefore, we suggest that the critical twist required for this flux rope to become kink unstable was not reached.

The writhes of the three axes were 0.29, 0.09 and -0.07 turns.
The axis that resulted in the largest value of average twist was the one with the smallest writhe, and vice versa.
Therefore, the sum of the twist and writhe is closer to being independent from the choice of the plane used to determine the axis (as expected; \citealt{torok2010writhe}).

\citet{guo2017magnetic} conclude that a good proxy of the magnetic helicity in the current-carrying field inside a finite volume is
 \begin{equation}
 H_{TW} = \Phi^{2}(Twist + Writhe),
 \label{eqn:H}
 \end{equation}
where $\Phi$ is magnetic flux.
In our case, using the total magnetic flux in the flux rope, this method gives a maximum helicity estimate of $\approx\num{3e41}$ Mx$^{2}$.
On the other hand, using the extrapolated field, the helicity of the closed, current-carrying field, H$_{J}$, in the extrapolation volume is $\approx$ \num{4e42} Mx$^{2}$, which is 14 times larger than H$_{TW}$.
Part of this discrepancy may originate from underestimating the magnetic flux in the flux rope when defining the boundary of the flux rope using the force-free parameter, $\alpha$.
However, since H$_{TW}$ scales with $\Phi^{2}$, there would need to be a factor of $\sqrt{14}$$\approx$3.7 error in the magnetic flux to fully explain the difference.

The mutual helicity between the potential field and current-carrying field, H$_{PJ}$ (as defined, \textit{e.g.}, by Equation 11 in \citealt{pariat2017relative}) $\approx$ \num{1e43} Mx$^{2}$.
The total helicity in the extrapolation volume is then given by H = H$_{J}$ + 2H$_{PJ}$ $\approx \num{2.4e43}$} Mx$^{2}$.
We are able to provide the first estimation of the eruptivity proxy recently introduced by \citet{pariat2017relative} for a NLFFF extrapolation of an AR, equal to $H_{J}/H=0.17$.

\subsection{Decay Index} \label{sec:decay}

 \begin{figure*}[!htb]
 \centerline{\includegraphics[width=1.0\textwidth,clip=]{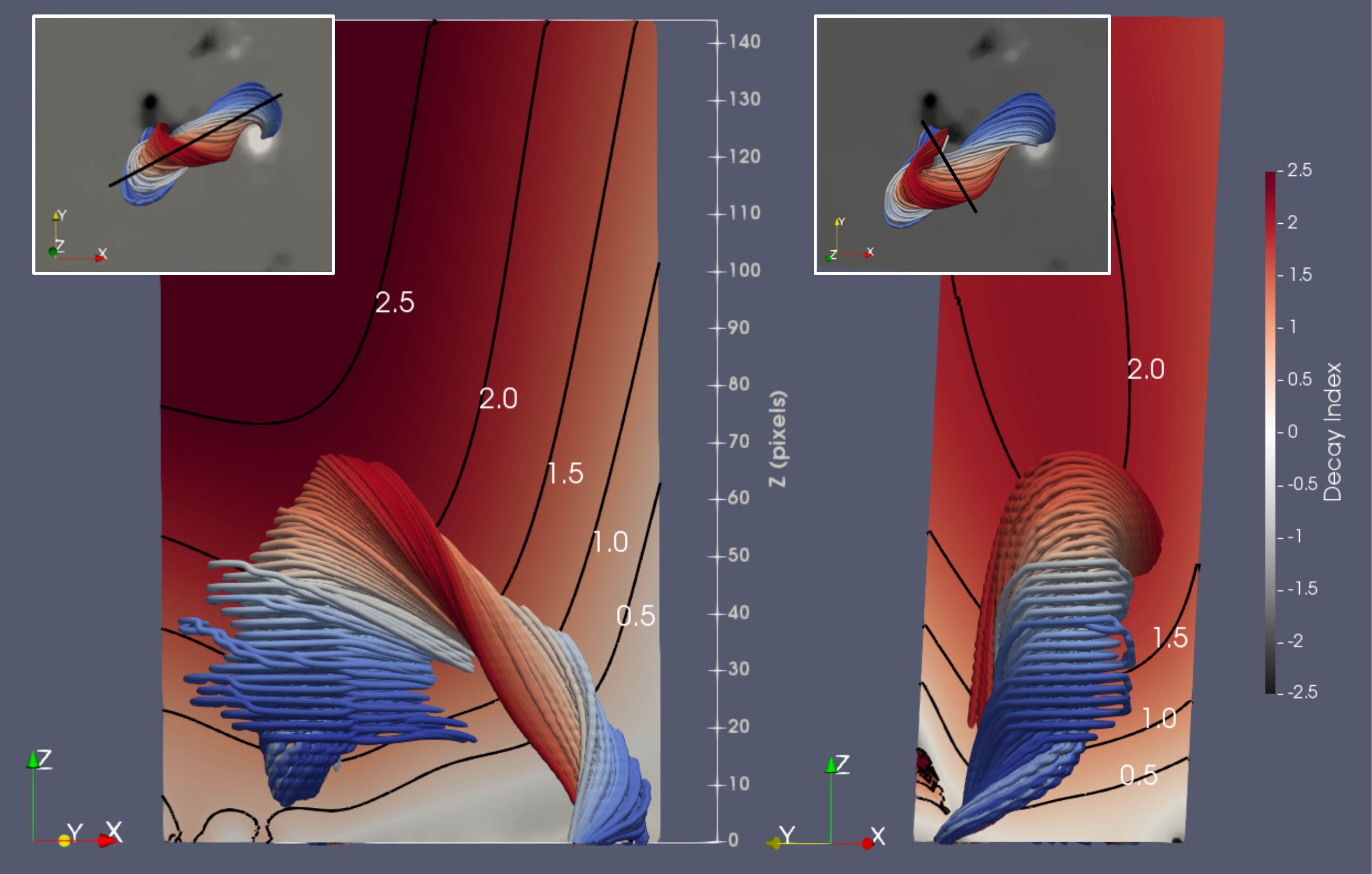}}
 \caption{The decay index is computed in slices along and perpendicular to the flux rope axis. The decay index at the centre of the flux rope is $\approx$1.8.}
 \label{fig:decay_index}
 \end{figure*}

A toroidal magnetic structure, such as a flux rope, may be unstable to perturbations if the overlying magnetic field strength decreases too rapidly with height. 
This phenomenon is referred to as the torus instability \citep{kliem2006torus}. 
The rate of change of magnetic field strength with height may be quantified by the decay index, $n$, defined as
 \begin{equation}
 n = - \frac{d \ln{B_{ext,p}}}{d \ln{R}},
 \label{eqn:decay_index}
 \end{equation}
where $B_{ext,p}$ is the strength of the poloidal component of the magnetic field external to the flux rope (non current-carrying), and $R$ is the major radius of the torus.

In order to estimate the decay index at the height of the flux rope, we perform a potential field extrapolation using the method of \citet{alissandrakis1981alpha} to approximate magnetic field external to the flux rope, and take the poloidal component as perpendicular to the central section of the flux rope axis.
We compute the decay index using the gradient of the poloidal field component in two planes inclined parallel with, and perpendicular to, the axis of the flux rope (see Figure \ref{fig:decay_index}).

The decay index near the centre (axis apex) of the extrapolated flux rope is equal to 1.8 (2.0). 
The value of the critical decay index required for the torus instability of a symmetric torus with a large aspect ratio is $3/2$ \citep{bateman1978instabilities,kliem2006torus}, and a number of studies have found similar values in magnetohydrodynamic simulations \citep{torok2007numerical,aulanier2010torus,kliem2013magnetohydrodynamic,zuccarello2015onset,zuccarello2016apparent}.
However, some studies have also found lower and higher critical decay indices (\textit{e.g.}, n$_{crit}$=1.1--1.3; \citealt{demoulin2010criteria}, and n$_{crit}$$\approx$2; \citealt{fan2010eruption}).
The extrapolated asymmetric flux rope lays in the upper range of the known stability limit for the torus instability, which is compatible with the eruption of the flux rope occurring one hour later.

\section{Conclusions} \label{sec:conclusion}

In this study, we test the hypothesis of \citeauthor{james2017on-disc} (\citeyear{james2017on-disc}; Paper I) that a magnetic flux rope formed in the corona of NOAA Active Region 11504 before erupting as a CME.
We produce a NLFFF extrapolation of the coronal magnetic field from a photospheric magnetogram one hour before the onset of eruption that closely supports the observational conclusions of Paper I. 


The axis of the extrapolated flux rope reaches $\approx$120 Mm above the photosphere ($\approx$150 Mm at the top of the flux rope). 
The decay index near the centre of the flux rope is $\approx$1.8, which is comparable to the critical value for the torus instability onset determined in other works. 
Therefore, we argue that the torus instability drove the eruption.

The extrapolation represents the coronal magnetic field one hour before the eruption. 
During that hour, it is likely that a triggering process further evolved the field (for reference, the coronal transit time in our model is about 140 seconds).
\citet{vemareddy2017prominence} observe the eruption of a highly twisted prominence (flux rope; 2.96 turns), and conclude that the helical kink instability caused the system to rise to the point at which the torus instability set in.
Our flux rope has an average twist in the range of 1.35--1.88 turns, and although this is similar to estimates of the critical twist for the kink instability to occur, the observations presented in Paper I suggest that kinking did not occur.
However, the flux rope in this study forms already high in the corona and does not need to be raised by another mechanism.
Instead, the motion of umbral sunspot fragments around each other may have inflated and therefore weakened overlying field, leading to the torus instability. 
Our NLFFF extrapolation closely matches this scenario proposed in Paper I, and therefore further supports the proposal that the eruption was triggered by photospheric flows and driven by the torus instability.


\acknowledgments 
AWJ, LMG, GV and LvDG acknowledge the support of the Leverhulme Trust Research Project Grant 2014-051. 
LMG also thanks the Royal Society for funding through their URF scheme.
YL was supported by NASA Contract NAS5-02139 (HMI) to Stanford University. 
M.C.M.C. acknowledges support by NASA contract NNG04EA00C (SDO/AIA) and grant NNX14AI14G (Heliophysics Grand Challenges Research).
YG is supported by NSFC 11773016, 11533005, and the fundamental research funds for the central universities 14380011.
LvDG is funded under STFC consolidated grant number ST/N000722/1 and also acknowledges the Hungarian Research grant OTKA K-109276. 
We thank the anonymous referee for their useful feedback.
Data are courtesy of the SDO science team. HMI and AIA are instruments onboard SDO, a mission for NASA's Living with a Star program.


\bibliographystyle{aasjournal} 

\end{document}